# Carbon Nanoarch Encapsulating Fe Nanowire on Ni (111)

Melanie David, Tomoya Kishi, Masanori Kisaku, Hiroshi Nakanishi, Hideaki Kasai

*Department of Applied Physics, Osaka University, Suita, Osaka 565-0871, Japan*

We investigate the stable structures of Fe-filled single-walled carbon nanotubes (SWNTs) on Ni(111), using density functional theory calculations. We find stable geometries and electronic states for the nanotube on Ni(111). We propose the possibility that the C-C bonds of carbon nanotube are broken by Fe wire and Ni surface. That is, when Fe-filled SWNT(3, 3) adsorb on Ni(111) surface, SWNT transforms into arch-like structure.

KEYWORDS: single-walled carbon nanotube, density functional theory, iron, nanowire, nanospintronics, nickel

## 1. Introduction

In recent years, nanospintronics has been the attention of many scientific studies. Nanospintronics is based on the ability to detect and control the spin degree-of-freedom of the charge carriers in nano-scale magnetic materials.[1] In particular, Fe nanowire has promising applications for spintronics devices such as conducting electron spin current or wiring the quantum dots. We have investigated Fe nanowires in isolated case (vacuum) and when placed on the Cu(111) surface.[2-4] Such Fe nanowires are protected from oxidation and the structure stabilizes when encapsulated within carbon nanotubes (CNT). The CNTs with ferromagnetic contacts exhibit spin dependent transport with high magneto-resistance effects.[5] When filled with ferromagnets, these CNTs demonstrate very high potential in providing the required magnetic properties, low dimensionality, and small volume for future nanoscale devices. Experimentally, CNTs filled with transition metal (Fe, Co, Ni) have been synthesized by chemical vapor deposition (CVD) using ferrocene as a precursor.[6]

In our previous theoretical studies, single-walled carbon nanotubes (SWNTs) encapsulating Fe nanowires have been examined.[7] We found that electronic and magnetic properties of those SWNTs depend on diameter of the nanotubes. In addition, the studies for SWNTs on various solid surfaces show that when the SWNT (6, 6) is absorbed on Si (001), its density of states around the Fermi level increases hence the metallic properties of the SWNT (6, 6) is enhanced.[8-9] In this way, the properties of SWNTs change when adsorbed on solid surfaces. Therefore, it is necessary to investigate the behavior of SWNT on solid surfaces, in order to apply the encapsulating transition atoms, such as Fe, for actual devices. These transition metals, Fe, Co and Ni, are used as catalysts for synthesizing SWNTs.[10] A recent study by Du *et. al.* shows that it has become possible to grow SWNTs on Ni substrates directly.[11]

In this paper, we investigate the stable structure and magnetic properties of SWNTs encapsulating Fe wires on Ni(111) surface using the density functional theory. We discuss their properties according to different adsorption sites of Fe nanowire.

**2. Model and calculations**

In this study, we perform spin-polarized first principles calculations using plane-waves and pseudopotentials.[12] We use generalized gradient approximation (GGA) for the exchange-correlation energy. The electron-ion interaction is described by optimized ultrasoft pseudopotentials using a cutoff energy of 35 Ry to limit the plane-wave basis set. Our system is consist of (3, 3) SWNT encapsulating Fe wire and a Ni(111) which is represented by three close-packed layers.   The top two layers of Ni(111), the Fe wire and (3, 3) SWNT are allowed to relax. We consider a vacuum region of about 14 Å separating the image surfaces. The two-dimensional Brillouin zone is sampled by 16 k-points. The initial magnetic moments are 2.2 $\mu_B$ and 0.6 $\mu_B$ for Fe atom and Ni atom, respectively.

For the super-cell approximation with periodic boundary conditions, we have one 1 Fe atom, 12 C atoms and 12 Ni atoms as shown in Fig. 1.  The (3, 3) armchair SWNTs on Ni (111) surface system has a cell length of 2.49 Å in the nanotube axis direction. We performed structural optimizations by minimizing the total energy until the residual forces become smaller than 0.05 eV/Å.

## 3. Results and discussion

The stable structures of the primitive (3, 3) SWNT adsorbed on Ni(111) and the (3,3) SWNT encapsulating Fe wire on Ni(111) are shown in Figs 2(a), 2(b) and 2(c). The structures of the SWNT and Ni surface do not change from those before adsorbing as shown in Fig 2(a). However, in the case of Fe wire enclosed in the SWNT, there is a significant difference in the structural change for each adsorption sites of Fe wire. When the Fe wire is relatively far from the Ni surface, the nearest Fe-C distance is 1.92Å. This is almost the same result obtained in the isolated (3, 3) SWNT encapsulating Fe wire wherein the Fe-C distance is 1.91 Å.[7] This indicates that there are no changes in the Fe-C bonding state even if the SWNT is adsorbed on Ni(111). In contrast, when Fe wire adsorbs on the nanotube's wall near the Ni surface, the SWNT transforms into arch-like structure (or a carbon nanoarch) with nearest neighbor distance of 2.08Å between Fe and C atoms as shown in Fig 2(c). This is 0.16 Å longer than the Fe-C distance when the Fe wire adsorbs on the nanotube's wall far from the Ni surface. Moreover, the Ni surface structure becomes slightly corrugated.

The electron charge density distributions in the cross section perpendicular to SWNT axis are shown in Figs. 3(a) and 3(b). Fig. 3(a) illustrates that when Fe wire adsorbs on the nanotube's wall far from the Ni surface, the charge density is about 0.6 electrons/Å$^3$ between Fe and C. Moreover, the bond is so strong that Fe wires have no magnetic moment. On the other hand, when Fe wire adsorbs on the nanotube's wall near the Ni surface side, the charge density is about 0.4 electrons/ Å$^3$ between Fe and C and the bond is weaker. As a result of this, SWNT have large magnetic moment (2.5 $\mu_B$ / Fe atom) hence it becomes a ferromagnetic metal. The total energy of the system when Fe wire adsorbs on the nanotube's wall near the Ni surface is 3.8 eV lower than the case when the Fe wire is far from the surface. We find that when the Fe wire adsorbs on the nanotube's wall near the Ni surface it becomes energetically stable thus a carbon nanoarch is formed.

We discuss the origin of this carbon nanoarch formation as follows. As shown in Fig. 2(c), the C atom which binds with Ni atom and Fe atom, also adheres to another two C atoms. This is almost tetrahedral structure with C in the center, and an sp$^3$-like orbital is formed. The electrons, which are originally C-C bonds, transfer to Fe-C bonds and Ni-C bonds. This leads to the breaking of C-C bonds, and the enhanced Ni-C bonds cause the corrugation on the Ni surface. Finally, the arch-like structure as shown in Fig 2(c) is formed by the Fe wire and Ni surface which act as catalysts for the breaking of C-C bonds.

## 4. Conclusions

We have investigated the stable structures of Fe-filled single-walled carbon nanotubes (SWNTs) on Ni(111), using density functional theory calculations. We find stable geometries and electronic states for the carbon nanotube on Ni(111). Fe-filled SWNTs transform themselves from tube to arch when placed near the Ni(111). We propose that C-C bond of carbon nanotube is broken by Fe wire and Ni surface. We find that the stable structure and magnetic moment of SWNTs encapsulating Fe wire adsorbed on Ni(111) significantly depend on the adsorption site and the position of Fe wire in the SWNT. Furthermore, Fe wire and Ni surface can act as catalysts for breaking the C-C bonds.

**Acknowledgments**

This work is supported by the Ministry of Education, Culture, Sports, Science and Technology of Japan (MEXT), through their Special Coordination Funds for the 21st Century Center of Excellence (COE) Program (G18) "Core Research and Advanced Education Center for Materials Science and Nano-Engineering" and Grants-in-Aid for Scientific Research (16510075) programs supported by the Japan Society for the Promotion of Science (JSPS), and the New Energy and Industrial Technology Development Organization (NEDO), through their Materials and Nanotechnology program. One of the authors (TK) acknowledges the support by research fellowships of Japan Society for the Promotion of Science for Young Scientists. Some of calculations have been done using the computer facilities of the ISSP Super Computer Center (University of Tokyo), the Yukawa Institute (Kyoto University), and the Japan Atomic Energy Research Institute (ITBL, JAERI).

**Figure Captions**

Fig. 1 Top view of the Ni(111) surface with the rectangle as the supercell. The size of the supercell is 2.5 Å in x axis, 8.6 Å in y axis and 22.4 Å in z axis.

Fig. 2 The optimized structures of (a)pure (3, 3) SWNT (b) (3, 3)SWNT encapsulating Fe wire adhered on the nanotube's wall far from Ni(111), and (c) (3, 3)SWNT in which Fe wire is near the Ni(111).

Fig. 3 Electron charge density distribution corresponding to a plane perpendicular to the nanotube axis when Fe wire adsorbs on the nanotube's wall (a) far from Ni surface and (b) near the Ni surface. The contour line spacing is a 0.2 eV/ Å$^3$.

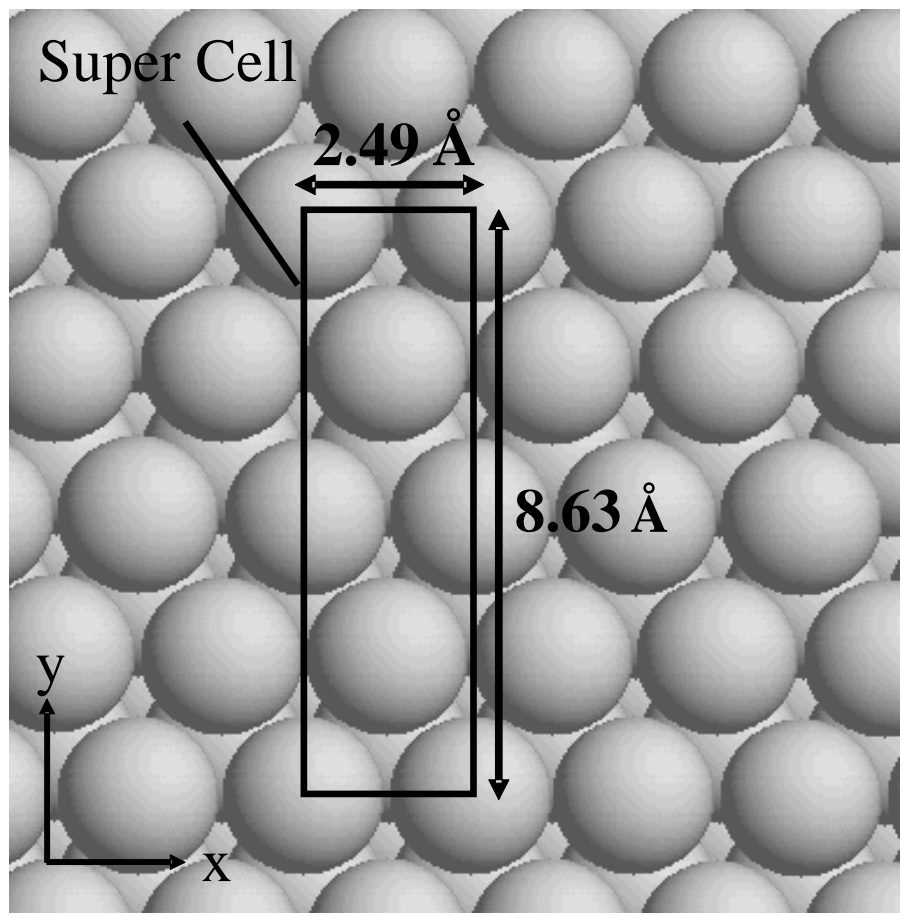

**Fig. 1**

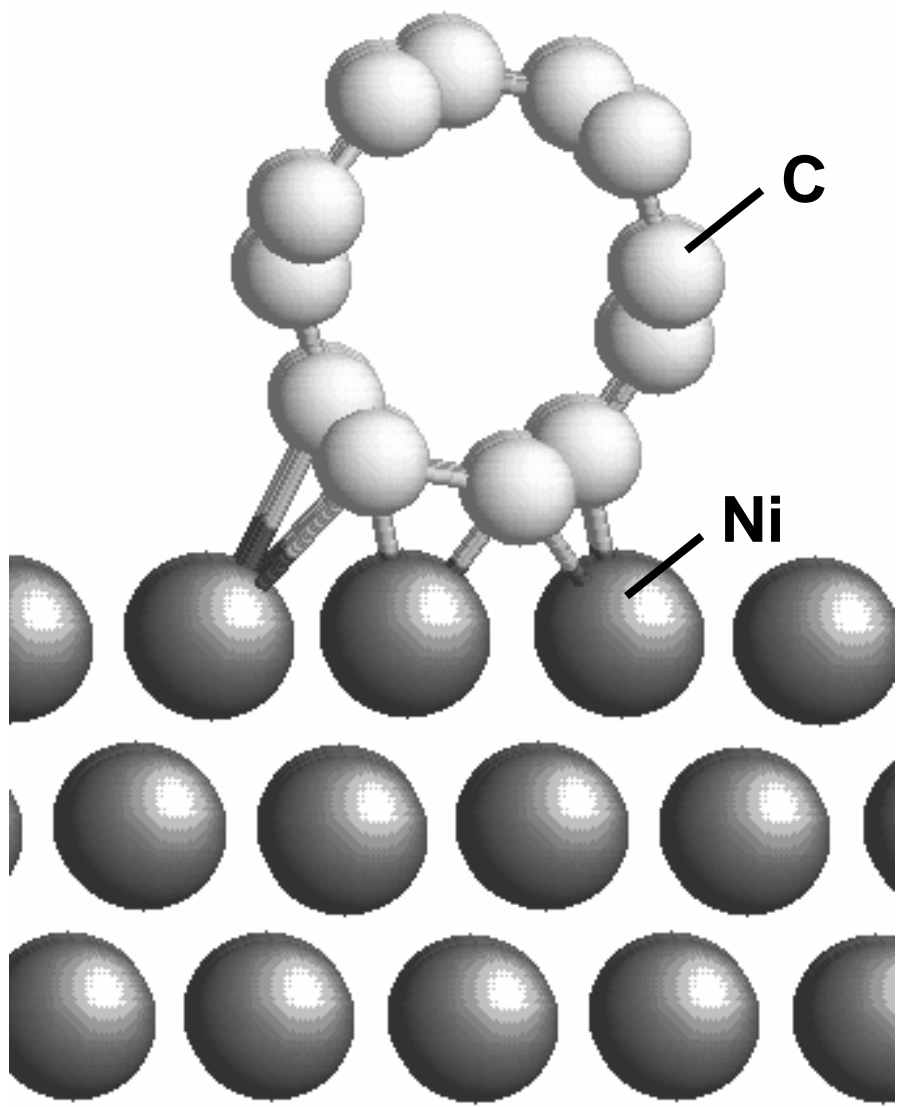

**Fig. 2(a)**

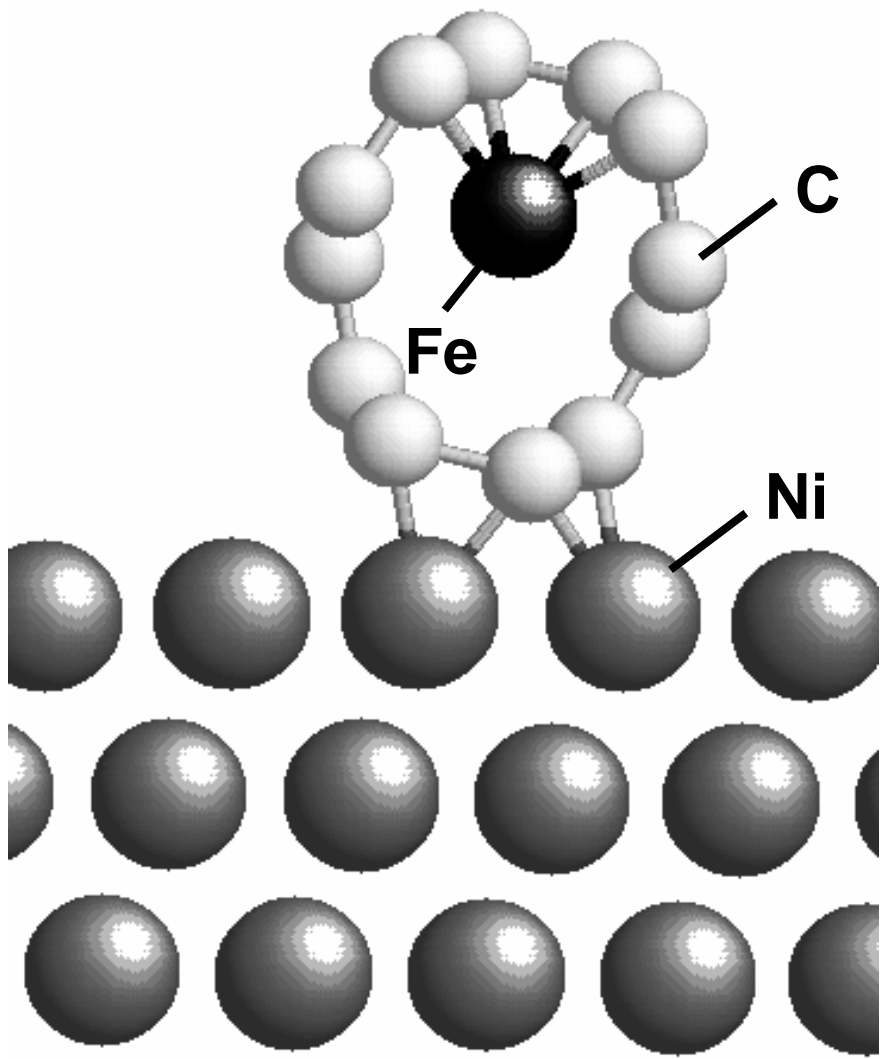

**Fig. 2(b)**

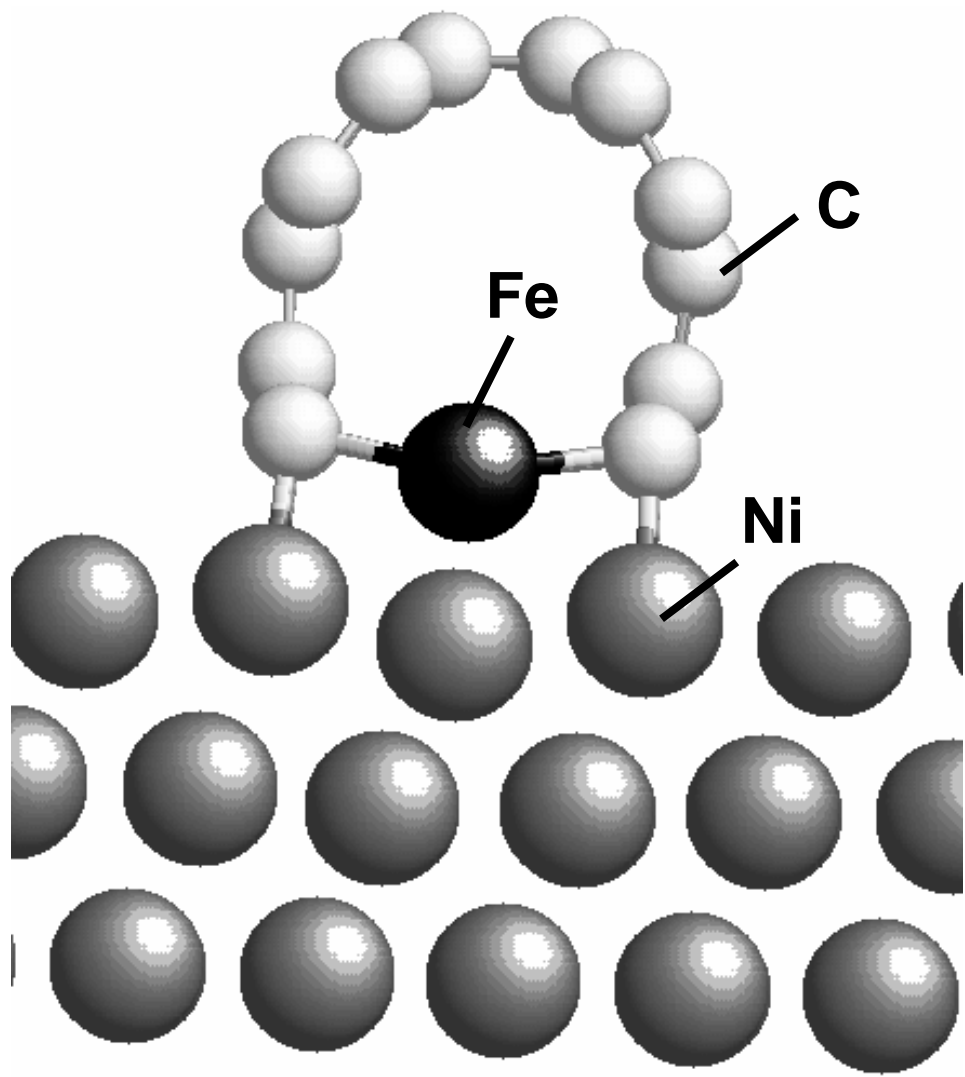

**Fig. 2(c)**

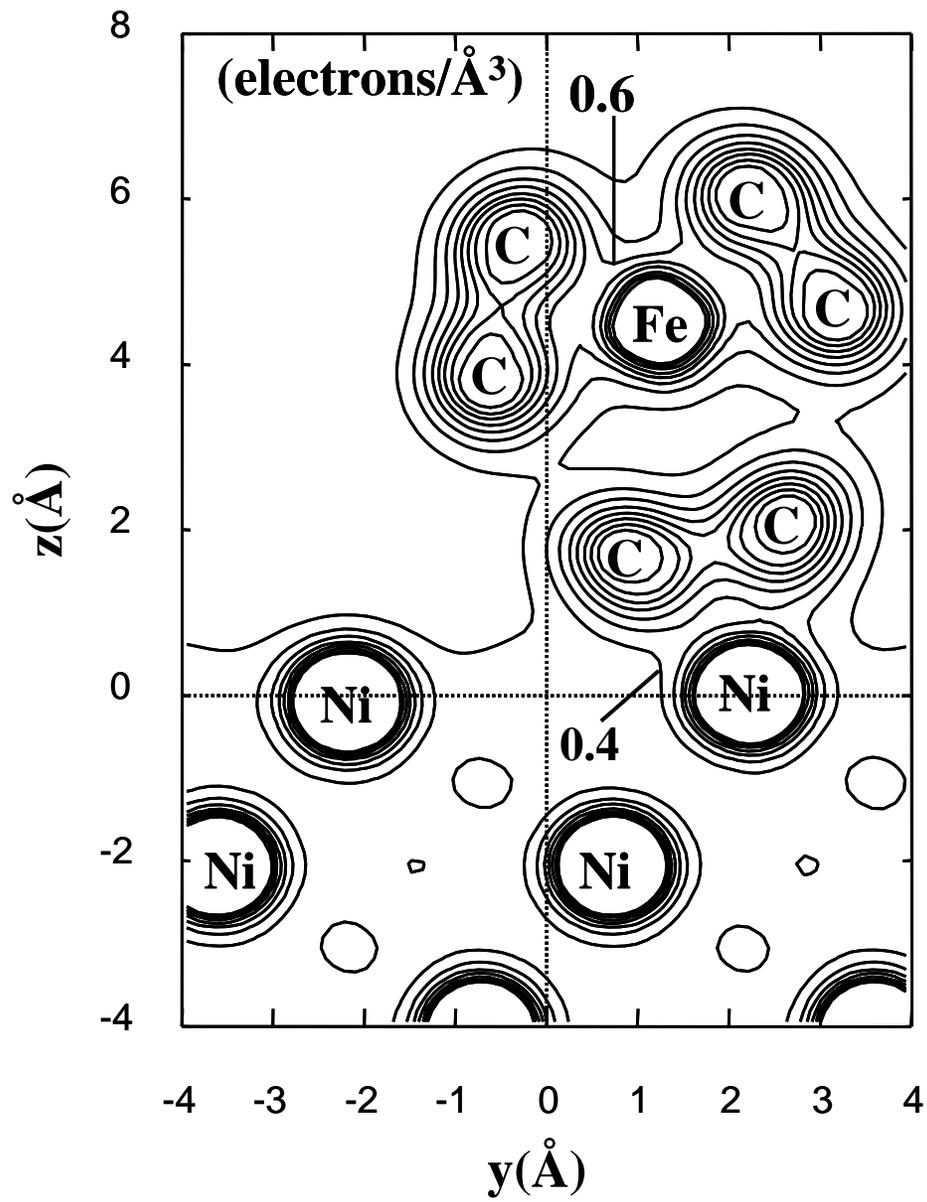

**Fig. 3(a)**

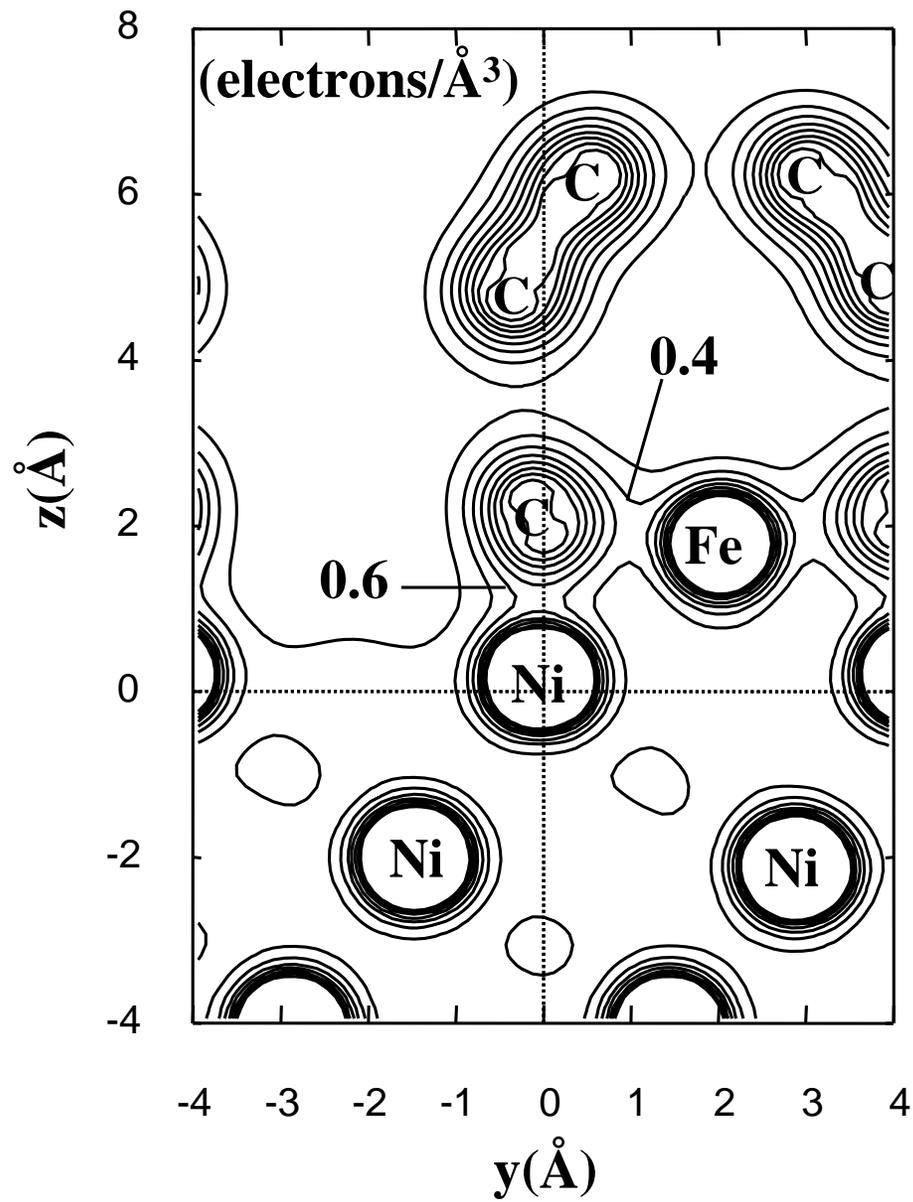

**Fig. 3(b)**